# DevRank: Mining Influential Developers In Github

Zhifang Liao1    Haozhi Jin1    Yifan Li1    Benhong Zhao1    Jinsong Wu2    Liu Shengzong3

(1) Department of Software Engineering,Central South University,China
(2) Department of Electrical Engineering, Universidad de Chile,Santiago
(3) Hunan University of Finance and Economics,China

*Abstract*—As the social coding is becoming increasingly popular, understanding the influence of developers can benefit various applications, such as advertisement for new projects and innovations. However, most existing works have focused only on ranking influential nodes in non-weighted and homogeneous networks, which are not able to transfer proper importance scores to the real important node. To rank developers in Github, we define developer's influence on the capacity of attracting attention which can be measured by the number of followers obtained in the future. We further defined a new method, DevRank, which ranks the developers by influence propagation through heterogeneous network constructed according to user behaviors, including "commit" and "follow". Our experiment compares the performance between DevRank and some other link analysis algorithms, the results have shown that DevRank can improve the ranking accuracy.

## Ⅰ. Introduction

Influence is a complicated force that affects the behaviors of people. It is well recognized that mining influential people are the keys of promotion. With the effect of "word of mouth", influential people can help to speed the promotion. As the emergence of social coding platforms, more and more developers construct their projects on the online platforms. Github[1] is the Facebook of social coding [9] and a popular online code hosting service built on top of Git, a decentralized version control system (DVCS)[10], which supports pull-based development paradigm[11]. As the biggest social coding platform, there are more than 1.7 million developers and 3 million projects on it, and it can be considered as a large-scale community[2]. Mining influential developers in Github can help to spread new information and innovation, it is also conducive to increase the efficiency of social coding. On traditional social networks like Facebook and Twitter，Internet sensations always have many "fans" on social website, and these sensations are easier to influence others due to their significant numbers of "fans", who are prone to attract new "fans". Similarly, influential developers and projects are also more attractive than others. In Github, developers can "follow" other developers and "star" some projects if they are interested in them. Thus, we define the influence of developers/projects as the ability to attract new "follow"/"star" in the future.

PageRank[3] and HITS[4] are frequently-used when measuring the influence or importance of nodes in social network. While both of the approaches work well in ranking the most influential nodes via analyzing the structures in social networks. Both of the approaches measure the importance of nodes by average propagation according to the direct structure with non-weight links, which cannot transfer proper importance to the real important node[5], these methods are also vulnerable to cheating links due to a single kind of edges. To solve this problem, we propose a new approach DeveloperRank (DevRank) which can be applied to a heterogeneous network that has different kinds of edges according to the most common events in Github, "follow" and "commit". DevRank classifies the target heterogeneous network into a homogeneous network(using "follow" information) and a bipartite network(using "commit" information), and computes influence scores using an unbalanced propagation strategy between nodes during iterations.

The remainder of this paper is structured as follows. In Section 2, we review the related work. Section 3 describes our proposed model for ranking developers in Github. Then, we present the experiment result and evaluations in Section4. Finally in section 5, we conclude the future research direction.

## Ⅱ. Related Work

Mining influential people in a social network is a meaningful work during long term exploration. There are a number of traditional methods to measure the authority of nodes, including degree, closeness, betweenness[4]. Eigenvector centrality based approaches, which measure nodes' authority by the maximum eigenvalue of the adjacency matrix of network such as PageRank, HITS, Katz and their derivative methods, show better performance than traditional methods, and they are widely used for ranking influential nodes in social networks. E. Katz[2] called the influential persons as "opinion leaders", and C. Wang[3] found that opinion leaders have higher PageRank scores than others. Zhengqiu Yang[4] conducted a study in the graph structure, social attributes and random walk model, proposed an approach called SocialRank to measure the influence of individual. Zeng Wei[5] has studied social network influence maximization problem and the existing algorithms, proposed a measure called HGA algorithm, which combines with a heuristic candidate seeding algorithm based on hidden influence and an activation model based on floating influence, and the HGA scores is an effective indicator of social influence of nodes in a social network. Hassan Sayyadi[6] constructed a author-paper heterogeneous network, and classified it into a citation network and an authorship network, then he proposed a measure to rank the nodes by FutureRank scores, which is calculated with three parts, including the hub scores when calculating the importance of citation network with HITS approach, the PageRank scores when calculating the importance of authorship network with PageRank approach, and the time scores, which is also an effective indicator to predict citation in the future. Yi Li[7] proposed a measure called CommRank, which aims at calculating the social influence of communities, and he also proposed an approximation algorithm based on CommRank to solve a influence maximization problem. Thung[8] constructed the developers network and projects network, and applied PageRank algorithm to calculate the importance of nodes in the networks which measure the influence of developers or projects, he also analyzed how strong the relationship is between the developers and the projects. All the methods show an improved performance over traditional approaches.

### III. DevRank Model

Most existing works using only "follow" to measure developers' influence, such as [8][9]. As we mentioned before, single kind of non-weghted edges can not transfer proper importance to the real important nodes. Based on this situation, we present a model, DevRank, that using not only "follow" but also "commit" to measure the developers' influence. In this section, first some statistics results are presented to explain why "commit" information is used to calculate the developers' influence. Then we introduce the structure of Github network constructed by both "follow" and "commit", and finally present DevRank based on Github network.

#### A. Follower-Commit relationship and Star-Commit relationship

Github is a platform with transparency environments for open source-style development, and allows any developers to clone any public repositories and commit changes at will. User can "follow" other users whom they are interested in, and "star" or "watch" the projects which they want to join in, they can construct their projects on the online platform, and "commit" their codes to other projects online[12].

First, we have picked about 9680 developers and 1396 projects started before 2012 to explore the relationship between three kinds of behaviors. Figure 1(a) shows the distribution of followers and commits, x denotes the number of developers' commits, y is the number of developers' followers. And Figure 1(b) shows the distribution of stars and commits, x is the number of projects' commits, y is the number of projects' stars. The follower-commit relation is shown as that developers' followers increase with developer's commits. Similarly, the star-commit relation is that projects' stars increase with the projects' commits received. Thus, we made the assumption that more commits leads to more influence, in other words, more commits a developer refers to a project, more influence scores propagate from the developer to the project.

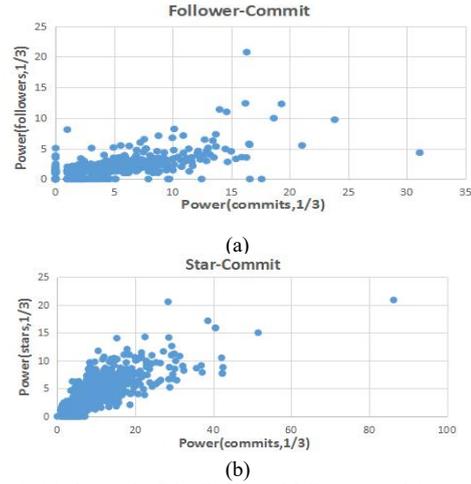

Figure 1: (a) shows the Distribution of followers and Commits (b) shows the Distribution of stars and Commits

To illustrate the assumption, we further conduct a research to confirm our assumption, we choose the same data set as before, and we count the average number of commits of developers and projects before 2012. We also count the average number of followers that developers obtained and the average number of stars that projects obtained during 2013. As shown in Figure 2, the red bars denote the number of followers newly obtained, and the yellow bars denote the number of stars newly obtained, and horizontal axis shows the commits range of developers or projects. For instance, developers whose commits range in (600,700] before 2012, obtained 80.43 followers during 2013 per person, projects whose commits range in (700,800] ,obtained 86.12 stars during 2013 per project. This diagram shows that, the number of new followers and new stars increases along with the increase of commits, which confirms the assumption that "commit" affects developers' influence scores and projects' influence scores positively, so nodes with more commits should gain more influence scores during iteration.

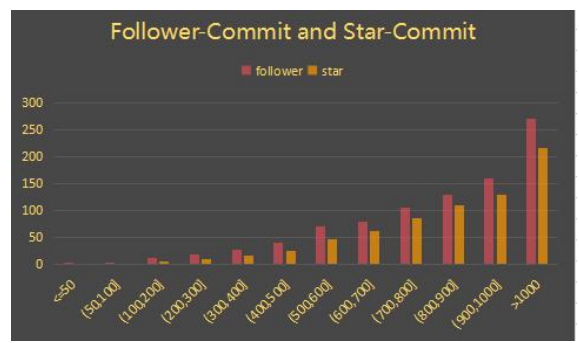

Figure 2: Average number of followers newly obtained based on the average number of commits of a developer(red bar),and average number of stars newly obtained based on the average number of commits of a project(yellow bar)

#### B. Decomposition of Github Network

As commit plays an important role in developers' influence, we extend the network by the data of "follow" and "commit" behaviors together. The target is to predict the number of future followers of the developers and the number of future stars of the projects in order to have a better ranking model. An abstract graph model is constructed as Figure 3. The network has two types of nodes, developers and projects. In addition, there are two types of edges, follow edges which are between developers, and commit edges which are between developers and projects, and they are both directed. The follow edges are non-weighted and the commit edges are weighted as the number of commits.

To analyze the nodes' influence by different behaviors, we partition the Github network into two networks to use two different kinds of behaviors in the same model which can be used to transfer influence through different networks. The first only contains the developer nodes and "follow" edges, so we can utilize PageRank to transfer influence scores between developers. The second network contains developers and projects, and they are connected by "commit" edges. The second network is a bipartite network which can mapped onto a HITS-type network. Thus, influence scores can be passed between developers and projects by HITS. Figure 4 shows the mapping.

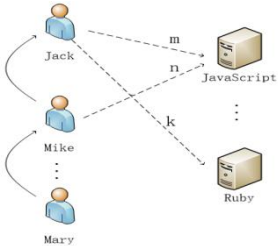
Figure 3: An example of the Github network

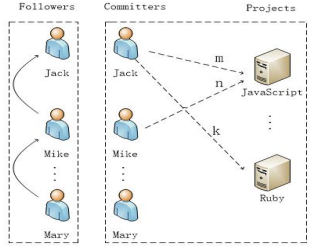
Figure 4: Network Decomposition: a follow-network and a commit-network

We define the two networks with adjacency matrices as follow:

$$M_{i,j}^{F} = \begin{cases} 1 & if\ d_i\ follows\ d_j; \\ 0 & otherwise; \end{cases} \quad (3)$$

where D is the set of developers, $M^F$ is the $|D| \times |D|$ follow matrix. For a node do not follow others, we consider it an unhealthy node, and we set $M^F_{i,j} = 1$ for all j, which is always an approach to treat dangling nodes in PageRank.

In addition, we define the commit matrix as follow:

$$M_{i,j}^{C} = \begin{cases} 1 & if\ d_i\ commits\ to\ p_j \\ 0 & otherwise \end{cases} \quad (4)$$

where P is the set of projects, $M^C$ is the $|D| \times |P|$ commit matrix.

### C. Proposed Algorithm: DevRank

Since two networks share nodes, we cannot compute rankings for each of the networks individually with different models. Instead, we present DevRank with asymmetric propagation strategy, which operates on both network, passing influence scores back and forth between the networks. Thus, the ranking algorithm is an iterative algorithm which runs one step of PageRank in follow-network, one step of HITS with asymmetric propagation strategy in commit-network. It then repeats the above steps until convergence.

First, we introduce the asymmetric propagation strategy of passing influence. For example, as we can see from Figure 3, which *Jack* has committed *m* times to *JavaScript* and *k* times to *Ruby*, so authority scores *Jack* transfer to JavaScript should be *m\*Inf(Jack)/(m+k)* during each iteration, where *Inf(Jack)* denotes the authority scores of Jack. Similarly, JavaScript has received *m* commits from Jack and n commits from Mike, so the authority scores JavaScript transfers to Jack should be *m\*Inf(JavaScript)/(m+n)* during each iteration, *Inf(JavaScript)* denotes the authority scores of JavaScript. We use propagation matrices to represent the propagation process in iterations:

$$M_{i,j}^{DP} = \frac{M_{i,j}^{C} \times n_{i,j}^{c}}{\sum_{i=1}^{|D|} n_{i,j}^{c}} \qquad M_{i,j}^{PD} = \frac{M_{i,j}^{C\ T} \times n_{j,i}^{c}}{\sum_{j=1}^{|P|} n_{i,j}^{c}} \quad (5)$$

where $M^{DP}$ represent the propagation matrix from developers to projects, and $M^{PD}$ denotes the propagation matrix from projects to developers, $n^c_{i,j}$ denotes the number of commits $d_i$ commits to $p_j$.

We use vectors to store ranking scores, where $R^P$ stores scores of projects and $R^D$ stores scores of developers. To solve the initialization problem and speed the iteration, we define the initial developer scores as the *PageRank* scores in the follow-network, the process is described as follow:

1.  $R_0^D = (Inf(d_1), Inf(d_2),...Inf(d_n)) = (1/n,1/n,...,1/n)$
2.  **While**  err > threshold:
3.      $R_n^D = M^F \times R_{n-1}^D$;
4.      $R_n^D = Normalize(\ R_n^D) = R_n^D/sum(R_n^D)$;
5.      err = $\sum|Inf_n(d_i)-Inf_{n-1}(d_i)|$
6.  $R_0^D = R_n^D$
7.  **Return**  $R_0^D$

where $M^F$ is the adjacent matrix of follow network. We have taken two common behaviors, "follow" and "commit", into consideration via defining the propagation matrix, Thus DevRank is an iterative algorithm which repeat steps as follow:

$$R_n^P = M^{DP} \times R_{n-1}^D \quad (6)$$

$$R_n^D = \alpha \times M^{PD} \times R_{n-1}^P + \beta \times M^F \times R_{n-1}^D + \frac{1-\alpha-\beta}{n_d} \quad (7)$$

where $R^P$ stores influence scores of projects and $R^D$ stores influence scores of developers. $M^{DP}$ represents the propagation matrix from developers to projects, and $M^{PD}$ is the matrix from projects to developers. $M^{DP}$ is not the matrix transpose of $M^{PD}$ because of the asymmetric propagation between developers and projects, $n_d$ represents the number of developers. *n* represents the iteration times. Projects' influence scores and developers' scores are updated mutually.

The scores of developers and projects update after each iteration, and we normalize two vectors in order to limit the scores' distribution. We set a threshold for iterative process, and the whole process is described as follows:

1. **While** *err > threshold:*
2. $R_n^P = M^{DP} \times R_{n-1}^D$;
3. $R_n^D = \alpha \times M^F \times R_{n-1}^D + \beta \times M^{PD} \times R_{n-1}^P + (1-\alpha-\beta)/n_d$;
4. $err = \sum|Inf_n(d_i) - Inf_{n-1}(d_i)| + \sum|Inf_n(p_i) - Inf_{n-1}(p_i)|$
5. **Return** $R^D, R^P$

where $R^D$ denotes the vector that stores the developers' authority scores. $R^P$ denotes the vector that stores the projects' authority scores. The initial value of $R^D$ is $(1/n_d, 1/n_d, ..., 1/n_d)$ and similarly the initial value of $R^P$ is $(1/n_p, 1/n_p, ..., 1/n_p)$, which ensures influence scores are all between 0 and 1. The property will hold after iteration too, since the computation performs an authority propagation and sum of the weights, $\alpha+\beta+(1-\alpha-\beta)$, is equal to one. The time complexity is $O(k*(2n_d n_p + 2n_d^2 + n_p^2))$, where $n_d$ denotes the number of developers, $n_p$ denotes the number of projects, and k is the iteration times.

## IV. Evaluation

In this section, we describe our data set and evaluate our proposed DevRank method with other traditional methods, and evaluate according to several performance criteria.

### A. Data set

We have evaluated *DevRank* on a real MySql dataset of Github, which were downloaded from GHTorrent, a scalable, queriable, offline mirror of data offered through Github Rest API. The data set contains records of 499000 developers' behavior and 5602 projects' information, where we consider projects with the same name or forked from the same repository as the same project. The set includes 505522 records of "follow", 689000 records of "commit", 12841 records of "star", 60000 records of "comment on commit" and so on, which are based on the events that occurred on Github from 2006 to 2014.

### B. Evaluation Setup

For evaluation, we split the data set into two sets: the first set, contains 1047550 records of "follow", 115473 records of "commit" and 89402 records of "star" before 2012, including 1047550 developers and 1320 projects. These developers have at least one record of "follow", "commit" and "star", and these projects are committed by these developers before 2012 in our data set. The second set is evaluation data set, including records of these developers and projects in the first set from 2012 to 2013. There were few records crossing the two sets, for instance, A is a developer who registered before 2012, while he followed someone registered after 2012, or committed to some projects established after 2012. These records have been removed in our evaluation. As we mentioned in Section Ⅲ, influential developers should obtain more followers in the future. Similarly, influential projects should obtain more stars in the future. Thus, we apply the first set in our experiment to compute influence scores of developers and projects, and use the second set to evaluate the result.

### C. Ranking: Evaluation and Approaches

For evaluating the ranking, we use two approaches: 1) precision curve, and 2) the *Pearson's ranking correlation* between the rankings provided by DevRank and PageRank computed on the test data.

We compare *DevRank* with *PageRank*, *HITS*, and different version of *DevRank*:

**DevRank:** Our proposed model with all available information("follow" and "commit"), it is an link analysis algorithm which updates influence scores in the both commit-network and follow-network.

$$R_n^D = \alpha * M^F * R_{n-1}^D + \beta * M^{PD} * R_{n-1}^P + \frac{1-\alpha-\beta}{n_d}$$

$$R_n^P = M^{DP} * R_{n-1}^D$$

**PageRank:** It is a traditional link analysis algorithm, which could be considered as *DevRank* in follow network and non-weighted commit-network, and we set $\alpha = 0.85$, and random jump with probability of 0.15, n denotes the number of developers. We use *PageRank* to compute the nodes' scores in the follow-network($\alpha=0.85, \beta=0$).

$$R_n^D = \alpha * M^F * R_{n-1}^P + \frac{1-\alpha}{n_d}$$

$$R_n^P = M^C * R_{n-1}^D$$

**HITS:** It is a traditional method with two types of nodes, hub-node and authority-node. The developers are set as hub-node, and projects are set as authority-node. Different with *DevRank*, *HITS* updates influence scores in non-weighted commit-network and does not use "follow" information($\alpha=0.85, \beta=0$).

$$R_n^D = \alpha * M^{C^T} * R_{n-1}^P + \frac{1-\alpha}{n_d}$$

$$R_n^P = M^C * R_{n-1}^D$$

**DevRank in Follow-Network(DF)**: A variant of *DevRank*, which update developers' influence in follow-network and update projects' influence in commit-network($\beta=0$).

$$R_n^D = \alpha * M^F * R_{n-1}^D + \frac{1-\alpha}{n_d}$$

$$R_n^P = M^{PD} \times R_{n-1}^D$$

**DevRank in Commit-Network (DC)**: A variant of our *DevRank* which update both developers' influence and projects' influence in commit-network ($\alpha=0$).

$$R_n^D = \beta * M^{PD} \times R_{n-1}^P + \frac{1-\beta}{n_d}$$

$$R_n^P = M^{DP} \times R_{n-1}^D$$

These methods are evaluated on the following two aspects:

**Influential developers ranking**: As we defined developers' influence as the capacity of attracting new followers, several approaches are used to compute developers' ranking scores on the first data set, and the results are evaluated with the precision of prediction on the second data set. Influential developers according to the rankings should obtain more followers than other developers in the future. Thus, the precision is defined as follow:

$$\Pr ecision = \frac{|Ranking_{top-k} \cap Followers_{top-k}|}{k}$$

**Influential projects ranking**: Similar to the case of the influential developers ranking, influential projects according to the rankings should obtain more stars than other projects in the future.

$$\Pr ecision = \frac{|Ranking_{top-k} \cap Stars_{top-k}|}{k}$$

### D. Effect of Parameters on Precision

Then we investigate the sensitivity of the performance of *DevRank* to different settings for the parameters which weights the "commit"($\alpha$), "follow"($\beta$), and the evaluation is based on Top-50 influential developers prediction.

Figure 5 shows the precision of *DevRank* for different values of $\alpha$ and $\beta$. The horizontal-axis shows the value of $\alpha$ and the vertical axis shows the value of $\beta$. Since $\alpha + \beta$ is always less than 1, the top right triangle of the map is empty. The darker the color in the heatmap, the higher the precision. Figure 5 shows the possible configurations of *DevRank*. For example, the precision shown on each edge of the heatmap triangle show a single part of *DevRank*. Values on the horizontal axis obtained for $\beta = 0$, so it means the horizontal edge shows all possible configurations of *DevRank* in follow-network(*DF*), while the vertical edge shows all possible configurations of *DevRank* in commit-network (*DC*) obtained for $\alpha=0$, and the hypotenuse edge shows the precision of DevRank which $\alpha + \beta$ is equals to 1.

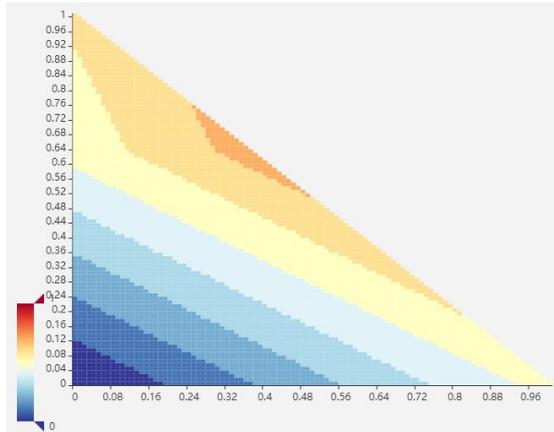

Figure 5: The precision of *DevRank* for different settings of the two weighting parameters, $\alpha$ (horizontal axis) and $\beta$ (vertical axis). In any point of the figure, $\alpha+\beta<=1$

The highest precision of *DevRank* is obtained at $\alpha=0.37$ and $\beta=0.63$, the corresponding precision is 0.74. We further oberserved that the precision of $DF(\alpha=1,\beta=0)$ is 0.54 and the precision of $DF(\alpha=0,\beta=1)$ is 0.62. And the precision decreases when $\alpha$ and $\beta$ decrease at the same time.

### E. Performance Evaluation

Next, various experiments are presented to evaluate the efficiency and effectiveness of *DevRank*.

*1) Influential developers prediction:* We begin this via exploring the precision of influential developers prediction for each algorithm. Figure 6 shows the precision of five methods.

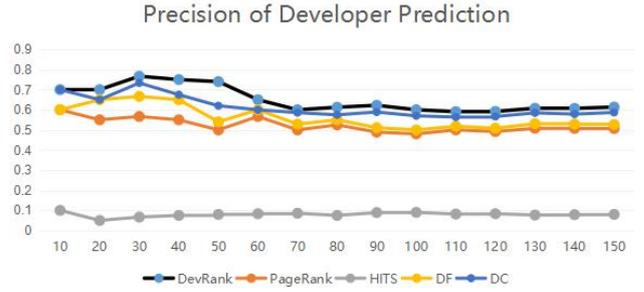

Figure 6: Precision in predicting Top-k influential developers

As we can see from the diagram, Figure 6 shows the precision of *DevRank* top k results for different values of k, *DevRank* provides significant improvement in ranking developers, the maximal precision of *DevRank* reaches 0.75 when we use it to predict top-30 influential developers, the precision of *DF* reaches 0.54 and the precision of *DC* reaches 0.62, and *DevRank*(using all available information) performs better than the both *DC*(using "commit" information) and *DF*(using "follow" information) at any point of the curve. In addition, we found that Precision(*DC*) ≥ Precision(*DF*) ≥ Precision(*PageRank*) is always true, since we use "commit" information to update developers' scores in *DC* and we use "follow" information to update projects' scores in *DF* and *PageRank*. Moreover, *DF* uses "commit" information to update projects' scores while *PageRank* does not, the comparison confirms that "commit" increased the prediction accuracy of influential developers, since developers' scores are propagated from the scores of projects during the iterations. Comparing our method with *HITS*, we can conclude that the propagation strategy has a remarkable influence in prediction precision, which means that considering the weight of edges does have a remarkable impact on prediction. Comparing *DevRank* with *PageRank*, we can observe that the *DevRank* preforms much better than the *PageRank*, which shows that "commit" obtains more influence scores than "follow" does.

*2)Correlation coefficient of rankings:* The correlation between followers obtained after 2012 and the rankings' scores of each method with Pearson correlation coefficient are shown in Figure 7. The highest correlation between the number of followers obtained in the future and *DevRank* ranking scores is close to 1, which is larger than the correlation of using both *HITS* and *PageRank*, and the correlations of *DevRank*, *DC* and *DF* are very close when k(the number of top-k developers) is less than 30, but the correlation of using *DF* goes down with k increases. This occurs, since both "commit" and "follow" information of the most influential developers are always more beneficial than that of others, since *DF*(only using "follow" information) has similar results to both *DevRank* and *DC*. This also confirms the precision of developer prediction and shows the significant improvement of accuracy via using *DevRank* in ranking influential developers than *PageRank* and *HITS*.

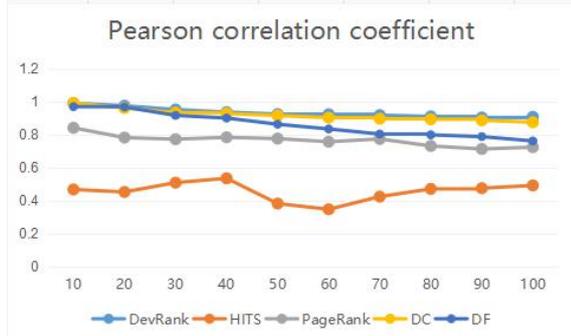

Figure 7：Correlation between the number of followers obtained in the future and the ranking's scores

We list the Top 10 influential developers of *DevRank* in Table 1, and we also show the most influential developers who obtain most followers in the second data set, which is the actual ranking.

Table 1: Top 10 developers obtain the most followers after 2012

| ID | Login | New followers after 2012 | Followers before 2012 | Commits before 2012 | PageRank | DevRank |
|---|---|---|---|---|---|---|
| 172799 | defunkt | 5567 | 6773 | 4201 | 1 | 1 |
| 81368 | addyosmani | 1573 | 3681 | 658 | 6 | 2 |
| 11886 | visionmedia | 1356 | 4598 | 14275 | 2 | 4 |
| 182902 | hakimel | 1132 | 2077 | 1072 | 18 | 3 |
| 92299 | mbostock | 1048 | 1976 | 1136 | 12 | 6 |
| 119755 | chriscoyier | 989 | 1320 | 354 | 34 | 5 |
| 106691 | mrdoob | 957 | 2514 | 1388 | 4 | 9 |
| 110998 | LeaVerou | 934 | 2113 | 601 | 9 | 13 |
| 51024 | lifesinger | 932 | 732 | 1542 | 153 | 8 |
| 164879 | mattt | 904 | 2377 | 617 | 5 | 10 |

Table 1 shows that most influential developers do commit a lot before 2012, but it does not mean the most influential one commits the most. For example, the developer whose ID number is 11886 commits the most, but he is not the most influential one. The number of commits can help one developer gain more authority scores while competing with others who commit to the same project, but committing to an influential is also important. Compared with *PageRank*, rankings of *DevRank* are closer to the actual rankings, which confirms the Pearson correlation coefficient between the number of followers obtained in the future and the ranking's scores in Figure 6.

*3) Influential projects prediction:* Now, we explore the precision of influential projects prediction. Figure 8 shows the precision of each method:

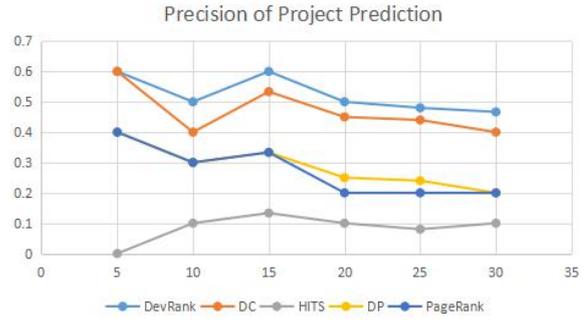

Figure 8: Precision in predicting Top-k influential projects

From this diagram, we can see that the accuracy using *DevRank* is also improved in project prediction compared with those for both *DC* and *DF*. *DF* still have similar performance with *PageRank*, and *HITS* still performs terrible in prediction due to the average propagation strategy, which confirms the effect of weight of edges on propagation. However, the prediction does not perform too well, the highest precision of *DevRank* is 0.6, since the development of projects are periodicaly changed[13]. For example, projects under development may gain more stars than the completed projects, and projects in the peak period will gain more attention than other periods.

*4) Convergence rate:* Then we evaluate the convergence rate of our algorithm, via setting different thresholds of iteration errors, we run our experiment on Intel(R) Core(TM) i5-4590 3.30GHZ CPU and 8GB memory. And the threshold of difference between the computed scores between two consecutive steps' scores was set to $e^{-8}$, $e^{-10}$ and $e^{-12}$ respectively.

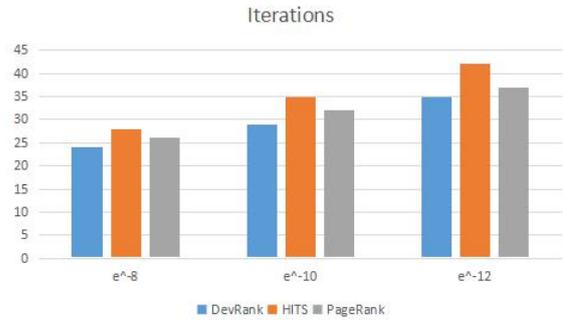

Figure 9: Iterations of different method

The results are shown in Figure 9. Comparing with *HITS* and *PageRank*, the number of iterations for convergence with *DevRank* is fewer. Hence, in addition to better precision and correlations, in terms of running time, *DevRank* is faster than *HITS* and *PageRank*.

F. *Effect of Time on Precision*

Next we evaluate the time effect on precision of influential developer prediciton of *DevRank* with three different time settings:

1). *Plan a*: we use the data from 2006-01-01 to 2012-12-31 as our training set and the data from 2013-01-01 to 2014-01-01 as our test set;

2). *Plan b*: we use the data from 2006-01-01 to 2011-12-31 as our training set and the data from 2012-01-01 to 2014-01-01 as our test set;

3). *Plan c*: we use the data from 2006-01-01 to 2010-01-01 as our training set and the data from 2011-01-01 to 2014-01-01 as our test set;

An experiment on top-k influential developers prediction with different settings is conducted and the results are shown in Figure 10. Similarly, developers with higher influence scores computed by *DevRank* in the training set should gain more new followers in the test set. The accuracy of prediction decrease as the training set gets smaller. That happens not only because of lack of training, but also due to timeliness of developers' influence. For example, influential developers may be normal after many years. This also happens frequently in the entertainment. Hence, the influence that our model aims to track is time-sensitive.

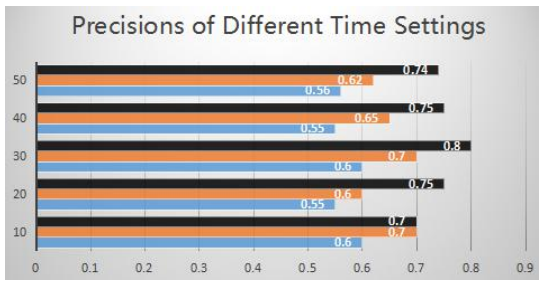

Figure 10: Precision of different time settings

V. Conclusion and Future Work

In this paper, we have proposed *DevRank* algorithm which utilizes the information of different user behaviors in Github, including "commit" and "follow", to rank the most influential developers effectively. While the influence of a developer can be measure by the capacity to attract followers, the number of followers that the developer will obtain measures how useful the developer has done to spread his influence. Our experiment evaluations have shown the precision of *DevRank*, which achieves significant improvement over that using other link analysis algorithms. The result of prediction precision showed that using the commits information significantly outperforms the tradition *PageRank* and *HITS*.

For future work, we plan to explore how time information of user behaviors affects the rankings. We also plan to investigate the influence that other behaviors contribute to, such as issue and comment. Furthermore, the rankings represent the global influence of developers in the network, we plan to investigate the local influence between developers.